\documentclass[aps,prd,twocolumn,twoside,nolongbibliography,nofootinbib, notitlepage, preprintnumbers,
showkeys,byrevtex,floatfix,amsmath,amssymb]
{revtex4-2}
\usepackage{marginnote,xparse,changepage,dcolumn,amsthm,amsfonts, bm, braket, array,amscd,amsmath,amssymb,graphicx,mathtools,multirow,placeins,slashed,verbatim,subfig,tensor,dsfont}
\usepackage[bookmarksnumbered,bookmarksopen=true,bookmarksopenlevel=1,pdfstartview=FitH, colorlinks=true, linkcolor=blue, citecolor=blue, urlcolor=blue]{hyperref}
\usepackage[all]{hypcap}
\usepackage{rotating}

\usepackage{soul}

\newcolumntype{M}[1]{>{$}{#1}<{$}}

\newcommand{\sst}[1]{{\scriptscriptstyle #1}}

\def\0{{\sst{(0)}}}
\def\1{{\sst{(1)}}}
\def\2{{\sst{(2)}}}
\def\3{{\sst{(3)}}}
\def\4{{\sst{(4)}}}
\def\5{{\sst{(5)}}}
\def\6{{\sst{(6)}}}
\def\7{{\sst{(7)}}}

\newcommand{\be}{\begin{equation}}
\newcommand{\ee}{\end{equation}}
\def\ba{\begin{array}}
\def\ea{\end{array}}

\newtheorem{claim}{Claim}

\newcommand{\R}{\mathds{R}}

\usepackage{soul}

\begin{document}

\title{Impossible measurements revisited}

\author{L. Borsten}
\email[]{lb199@hw.ac.uk}
\author{I. Jubb}
\email[]{ijubb@stp.dias.ie}
\author{G. Kells}
\email[]{gkells@stp.dias.ie}

\affiliation{School of Theoretical Physics, Dublin Institute for Advanced Studies,
10 Burlington Road, Dublin 4, Ireland}

\date{\today}

\begin{abstract}

It is by now well recognized that the naive application of the projection postulate on composite quantum systems can induce signalling between their constituent components, indicative of a breakdown of causality in a relativistic spacetime context. Here we introduce a necessary and sufficient condition for an ideal measurement of an observable to be nonsignalling. As well as being particularly simple, it generalizes previous no-signalling conditions in that it allows for degeneracies and can be applied to all bounded self-adjoint operators. The condition is used to establish that arbitrary sums of local observables will not signal, in accordance with our expectations from relativistic quantum field theory. On the other hand, it is shown that the measurement of the tensor product of commuting local observables, for example bipartite operators of the form $A\otimes B$, can in fact signal, contrary to the widely held belief that such measurements are always locally realizable. The implications for the notion of measurement in relativistic quantum field theory are addressed; it appears that the most straightforward application of the standard  quantum formalism generically leads to violations of causality. We conclude that either the class of observables that can be measured should be restricted and/or that the naive translation of the measurement framework of quantum theory, in particular the projection postulate, to quantum field theory should be reevaluated.

\end{abstract}


\preprint{DIAS-STP-19-10-XX}

\maketitle

\section{Introduction} The violation of the Bell inequality by entangled quantum states~\cite{Aspect:1982} rules out\footnote{To argue this one needs to make certain reasonable assumptions, which one could nonetheless question.} \emph{local} hidden variable models~\cite{Bell:1964kc}. In this sense quantum theory is nonlocal. However, as is well known, entanglement does not allow for instantaneous signalling; Beolagh's (receiver) marginal probabilities are independent of Aoife's (sender) measurement choices. Einstein's ``spooky action at a distance'' characterization notwithstanding, entanglement and causality are happy bedfellows. Yet, as has been long recognized~\cite{landau1931erweiterung,Dirac:1933,   Hellwig:1970st, PhysRevD.21.3316, PhysRevD.24.359, PhysRevD.34.1805, Sorkin:1993gg, Popescu:1993xc, Beckman:2001qs, Benincasa:2012rb, Jonsson:2013ikb, Martin-Martinez:2015psa,Tjoa:2018uvp}, the tension between quantum theory and causality persists for more elementary reasons. 

The conventional account of ideal measurement leads to superluminal signalling when straightforwardly applied, in the sense precisely captured in~\cite{Sorkin:1993gg}, to certain observables in spacetime~\cite{Sorkin:1993gg, Popescu:1993xc, Beckman:2001qs, Benincasa:2012rb}. Examples of such observables in relativistic quantum field theory (QFT) include one-particle wave-packet states~\cite{Sorkin:1993gg, Benincasa:2012rb}, Unruh-DeWitt detectors with nonlocal field couplings~\cite{Benincasa:2012rb} and Wilson loops in non-Abelian gauge theories~\cite{Beckman:2001ck}. Such causality violations imply that it is \emph{impossible} to perform an ideal measurement of one of these observables, within its associated spacetime region.

On the other hand, there are numerous examples of signalling in nonrelativistic quantum theory, incomplete Bell basis measurements~\cite{Sorkin:1993gg, Popescu:1993xc} for example. This does not mean that such observables cannot be ideally measured (they are routinely), only that the constituent systems must be brought into causal contact to realize the measurement. 

Let us elaborate for clarity.  In the  nonrelativistic setting we consider composite multiparty  systems (laboratories), that are strictly isolated in the sense that no signal may propagate from one component to another. This could be due to the fact that the laboratories are  spacelike separated, or it could be imposed by some other physical obstruction (or for the purposes of argument, simplify imposed by fiat).  This is precisely the situation for Bell-type experiments. To close the ``locality loophole'' one would like to ensure that no signal can pass between the laboratories, to be sure that  any violations of the Bell-type  inequalities are really due to entanglement and not  some hidden  signalling. In practice, this can indeed be achieved by making the laboratories spacelike separated \cite{giustina2015significant}.

Given such isolated joint systems, one can ask if there are joint observables whose ideal measurement could be used to induce a signal between the two isolated laboratories. As already emphasized, and as we shall show,  generalizations of Bell-type measurements (which are manifestly local) on the joint system cannot induce signalling. On the other hand, we shall show that other, seemingly innocuous, joint observables can induce signalling. This does not mean that they cannot be ideally measured, only that they must be brought into causal contact to do so. Indeed, experimental implementations of signalling measurements \emph{can} be performed, but if the constituents are separated it is always the case that the measurement will take at least the light travel time between them to complete. 

It is therefore meaningful, in relativistic or non-relativistic quantum theory, to discuss causal constraints on what observables may be ideally measured in a given spacetime region~\cite{Popescu:1993xc, Beckman:2001qs}. In particular, in~\cite{Beckman:2001qs} they consider bipartite systems with tensor product Hilbert spaces and give a comprehensive causal characterization of quantum channels, which includes ideal measurement as a special case, separating quantum operators into those that are (semi)localizable,  (semi)causal, and signalling.

The proper setting for studying causal constraints on measurements is  QFT, where causality is encoded through the (anti)commutation of field operators at spacelike separated points.  This ensures, for example, that correlation functions will always respect causality. It does not, however, protect us against signalling through the naive application of the projection postulate~\cite{Sorkin:1993gg}.  QFT is our very best framework for making predictions for the probabilities of the outcomes of measurements and there are certainly observables we \emph{can} measure, scattering amplitudes being the prime example. In this case however, the asymptotic nature of the S-matrix washes away all causal considerations. The question still remains, however, whether one can do ideal measurements of the many other observables in QFT; smeared field operators for example. Furthermore, if ideal measurements of certain observables are not possible, what sort of measurements are? Can one get arbitrarily close to an ideal measurement?

Here we revisit the conditions under which an ideal measurement will not signal. We begin by articulating an expression of the projection postulate robust enough to accommodate all observables, i.e.~arbitrary self-adjoint operators, and is therefore robust enough to be applied to the continuum setting of QFT, which differs in several meaningful ways from the discrete case\footnote{For example, in contrast to the discrete case, the total Hilbert space of a continuum QFT is not a tensor product of local Hilbert spaces associated to each point in space (see the \emph{split property} discussion in~\cite{fewster2019algebraic}).}. Our expression of the projection postulate relies on the notion of a \emph{measurement resolution} $\mathcal{B}$, a set of disjoint Borel subsets covering the real line. Physically, it can be interpreted as the ability for the measurement apparatus to distinguish the possible outcomes for a given observable. Then, a  general criterion for no-signalling is derived which can be applied to \emph{any} state update map, $\mathcal{E}$, including Projection Operator-Valued Measures (POVM's) and the case of interest here --- ideal measurements. We then apply the criterion to two simple cases.  First, it is used to rederive the result~\cite{Benincasa:2012rb} that sums of local compact self-adjoint operators cannot signal, which accords with our expectations from QFT and subsumes the Bell-type measurements. Second, it is used to show that the tensor product of local compact self-adjoint operators can indeed signal, and we give a simple two-qubit example using a separable state. This somewhat surprising result contradicts the claim that such observables are locally realizable and hence causal~\cite{Beckman:2001qs}. Indeed, such observables are routinely used in various quantum information theoretic contexts. Finally, the possible implications of these results are considered in the relativistic context of QFT.

It should be stressed that our no-signalling criterion is agnostic to the particular experimental setup or the measurement model adopted. This is because it is formulated in terms of the state update map, which is general enough to encompass any current description of quantum measurement. Our criterion should therefore be understood as a theoretical limit on what is possible, independent of how it is attempted. The evidence presented here indicates that this limit depends intimately on the precise form of the QFT observable under consideration.


\section{The setup} 

Consider the usual textbook notion of an ideal measurement of an observable: (i) the measurement outcomes are the eigenvalues of the self-adjoint operator corresponding to the observable; (ii)  the probability of a specific  outcome is given by the Born rule; (iii) the post-measurement state of the system is given by the  projection postulate.

Let us consider (i)-(iii) more carefully. Take some quantum system with Hilbert space $\mathcal{H}$.  We denote the algebra of operators $\mathcal{H}$ by  $\mathfrak{A}$. A compact self-adjoint operator $O$ has a discrete spectrum of eigenvalues and may be written as
\begin{equation}
O=\sum_n \lambda_n E_n \;\;\; ,
\end{equation}
where $\lambda_n$ are the  distinct  eigenvalues and $E_n$ are the   associated (not necessarily rank-1) projectors onto the corresponding eigenspaces,  which resolve the identity, i.e.
\begin{equation}
\sum_n E_n = \mathds{1} \;\;\; .
\end{equation}

Following a measurement of $O$, yielding outcome $\lambda_n$ on  an initial  state $\rho$, the  corresponding post-measurement density matrix, $\rho_n$, is given by the projection postulate:
\begin{equation}
\rho\mapsto \rho_n = \frac{1}{p_n}E_n \rho E_n \;\;\; ,
\end{equation}
where $p_n =\text{tr}(\rho E_n)$ is the probability of observing $\lambda_n$. If one conditions on the outcome $\lambda_n$, the state $\rho_n$ is used to calculate probabilities of any subsequent measurement. If one does not condition on any particular outcome of the initial measurement one must consider the distribution over all possible post-measurement states,  $\rho_n$,  weighted by their respective probabilities  $p_n$:
\begin{equation}
\rho\mapsto \rho' = \sum_n p_n \rho_n = \sum_n E_n \rho E_n\;\;\; .
\end{equation}
This is the projection postulate for a \emph{non-selective} measurement \cite{Hellwig:1970st}. 

The preceding account relies on the existence of a  discrete spectrum.   Nevertheless, for an arbitrary self-adjoint operator $O$ the spectral theorem states
\begin{equation}
O = \int_{-\infty}^{\infty}\lambda \, dE(\lambda) \;\;\; ,
\end{equation}
where $E(\cdot)$ is the projection-valued measure for $O$. That is, $E(\cdot)$ maps  Borel subsets $B\subseteq \mathds{R}$ to projectors on $\mathcal{H}$~\cite{reed1980functional}. This allows  the projection postulate to be articulated for arbitrary self-adjoint operators. Consider a set of mutually disjoint Borel sets $\mathcal{B} = \lbrace B_n \rbrace_{n\in\mathcal{I}}$ (where $\mathcal{I}$ is some countable indexing set) that covers $\mathds{R}$. For example,  $\mathcal{B} = \lbrace [n,n+1)\rbrace_{n\in\mathds{Z}}$. Physically, each $B_n$ corresponds to a possible bin that the measurement outcome can fall into, and in this way  $\mathcal{B}$ specifies the  resolution of the measurement apparatus. The corresponding projectors, $E_n := E(B_n)$, resolve the identity, i.e.
$
\sum_{n\in\mathcal{I}}E_n = \mathds{1}$.  Following a non-selective measurement of $O$, with resolution  $\mathcal{B}$, the projection postulate  is given by
\begin{equation}
\rho\mapsto \rho' =\mathcal{E}_{O,\mathcal{B}}(\rho) := \sum_{n\in\mathcal{I}}E_n \rho E_n  \;\;\; ,
\end{equation}
where we have defined the trace-preserving map $\mathcal{E}_{O,\mathcal{B}} \, : \, \mathfrak{A} \mapsto \mathfrak{A}$. The conditions  $\text{tr}(\rho)=1$ and $\text{tr}(\rho A^{\dagger}A)\geq 0$ for any $A\in\mathfrak{A}$ are preserved by $\mathcal{E}_{O,\mathcal{B}}$, so that $\rho'$ is a valid state.


\section{Signalling}

Let us now consider the straightforward application of ideal measurement to sequences of observables in a spacetime $\mathcal{M}$, following the framework introduced in~\cite{Sorkin:1993gg}. To each region of spacetime, $\mathcal{R}\subset\mathcal{M}$, there is a subalgebra of operators $\mathfrak{A}(\mathcal{R})\subseteq \mathfrak{A}$, such that $[\mathfrak{A}(\mathcal{R}),\mathfrak{A}(\mathcal{R}')]=0$ for $\mathcal{R}$ and $\mathcal{R}'$ mutually spacelike\footnote{Here we are working in the Heisenberg picture, where the operators carry the dynamics.}. As stated above, this is the usual way in which causality is encoded in QFT through (anti)commutation relations, but as we will see, it is not strong enough on its own to prevent superluminal signalling under ideal measurements.

\begin{figure}
\includegraphics[width=0.48\textwidth]{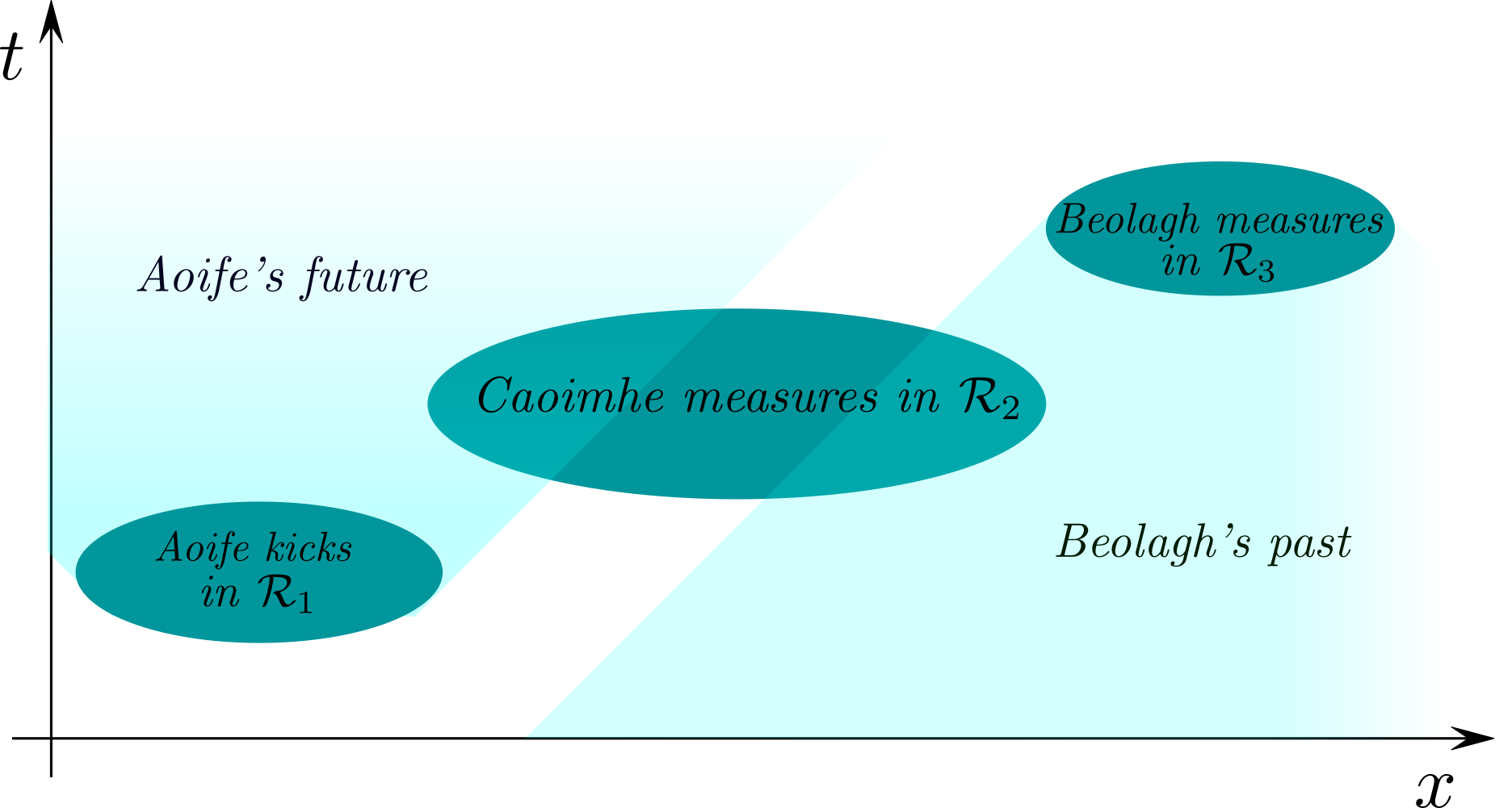}
\caption{Spacetime diagram of the protocol in the lab frame. Light rays are at $45^{\circ}$. Aoife kicks locally with $U_{\gamma}$ in the spacetime region $\mathcal{R}_1$, Caoimhe measures $O_2$ in $\mathcal{R}_2$, then Beolagh measures $O_3$ in $\mathcal{R}_3$. Aoife's kick can causally affect Caoimhe's measurement, as Caoimhe's region is partly in the future of Aoife's region. Caoimhe's measurement can similarly affect Beolagh's measurement, as it is partly in the past of Beolagh's region. Aoife's kick cannot causally affect Beolagh however, as their regions are spacelike; light from Aoife cannot reach Beolagh in time for his measurement.}
\label{fig:setup}
\end{figure}

We consider three parties whose actions on a shared quantum system are restricted to lie in three separate spacetime regions: Aoife in $\mathcal{R}_1$, Beolagh in $\mathcal{R}_3$, and Caoimhe in $\mathcal{R}_2$. The regions are such that every point in $\mathcal{R}_1$ is spacelike to every point in $\mathcal{R}_3$, while there are some points of $\mathcal{R}_2$ that are to the future (past) of some points in $\mathcal{R}_1$ ($\mathcal{R}_3$) (see figure~\ref{fig:setup}). This means that Aoife's actions in $\mathcal{R}_1$ \emph{can} affect Caoimhe in $\mathcal{R}_2$, and that Caoimhe's actions \emph{can} affect Beloagh in $\mathcal{R}_3$, but, that Aoife's actions \emph{cannot} influence Beolagh; to do so would violate causality. We  implicitly assume that there are many independent duplicates of the quantum system, such that each party can make multiple simultaneous measurements to build up statistics. 

The protocol, as seen from the lab frame, is as follows. The initial state of the shared quantum system is $\rho_0$.  
First, Aoife applies a local unitary ``kick'' $U_\gamma=e^{i\gamma O_1}$, for some self-adjoint $O_1\in\mathfrak{A}(\mathcal{R}_1)$, of ``strength'' $\gamma\in\R$:
\begin{equation}
\rho_0\mapsto \rho_\gamma=U_{\gamma}\rho_0U_\gamma^{\dagger} \;\;\; .
\end{equation}
Then, Caoimhe measures some self-adjoint $O_2\in\mathfrak{A}(\mathcal{R}_2)$ with measurement resolution $\mathcal{B}$. Since there is no communication between Caoimhe and the other parties, the state is updated as
\begin{equation}
\rho_\gamma\mapsto\rho_{\gamma}'=\mathcal{E}_{O_2,\mathcal{B}} (\rho_\gamma) \;\;\; .
\end{equation}
Finally, Beolagh measures the expectation value,
\begin{equation}
\langle O_3 \rangle = \text{tr}(\rho'_\gamma O_3) \;\;\; ,
\end{equation}
of some self-adjoint operator $O_3\in\mathfrak{A}(\mathcal{R}_3)$. Note that Caoimhe does not need to communicate the results of her measurement to Aoife and Beolagh. 

Recall that Aoife's local actions cannot influence Beolagh in our setup. If Beolagh can determine in \emph{any} way that Aoife has kicked the state, then causality is violated, thus implying that some part of the preceding protocol \emph{cannot} be implemented in a lab. Mathematically, a dependence on Aoife's actions in Beolagh's measurement amounts to a dependence on the kick strength $\gamma$ in Beolagh's expectation value $\langle O_3 \rangle$. That is, if $\langle O_3 \rangle$ is a function of $\gamma$ we have a \emph{superluminal} signal. Crucially, such a dependence can only be due to Caoimhe's measurement, as if she had not measured we would have $\rho_{\gamma}=\rho_{\gamma}'$, and hence $\langle O_3 \rangle = \text{tr}( \rho_{\gamma} O_3 ) = \text{tr}( \rho_0 O_3 )$, using the fact that $[U_{\gamma} , O_3 ]=0$. $\langle O_3 \rangle$ is clearly independent of $\gamma$ in this case. The above protocol then serves to test whether \emph{Caoimhe's} ideal measurement of $O_2$ is possible in the region $\mathcal{R}_2$.

It is worth noting that we are using spacetime causality here to separate Aoife and Beolagh as ultimately we have in mind measurements in QFT. However, we shall also address operators in non-relativistic quantum theory. In that case one could simply consider them as isolated laboratories -- Aoife and Beolagh only have access to their respective factors of the Hilbert space --  imposing that there be no signalling in the usual sense that Aoife's local actions cannot affect Beolagh's statistics.

For any given  $O_2$ and measurement resolution $\mathcal{B}$ one must check whether there is some initial state $\rho_0$ and observables $O_1, O_3$ for which there is a signal. One could argue that only the operators $O_2$ and resolutions $\mathcal{B}$ that do not signal can be ideally measured in reality. Alternatively, it could be taken as motivation to reassess the notion of  measurement in a relativistic context (see~\cite{Fewster:2018qbm,bostelmann2020impossible} for example, and further discussion below). Either way, it would be helpful to have a clear no-signalling condition:
\begin{claim}[The Causal Criterion]\label{claim1}
An operator $O_2 \in\mathfrak{A}(\mathcal{R}_2)$ with  resolution $\mathcal{B}$, will not enable signalling  iff $\, [\mathcal{E}_{O_2,\mathcal{B}}(O_3),O_1]=0$, as an operator equation, for all $O_{1,3}\in\mathfrak{A}(\mathcal{R}_{1,3})$.
\end{claim}

\noindent We  argue as follows. Using the cyclicality  of the trace and  $U_\gamma  = \sum_{n=0}^{\infty} \frac{(i\gamma)^n}{n!}O_1^n$, the expectation value   $\langle O_3\rangle$ can  be written 
\begin{equation}\label{eq:series_expansion_of_expected_value}
\langle O_3\rangle
=\sum_{n=0}^{\infty} \frac{(i\gamma)^n}{n!} \text{tr}\left( \rho_0 \, [ \mathcal{E}_{O_2,\mathcal{B}}(O_3),O_1]_n \right) \;\;\; ,
\end{equation}
where $[X,Y]_0 := X$ and $[X,Y]_{n+1} := [[X,Y],Y]_n$.
No signalling implies  $\langle O_3 \rangle$ is independent of $\gamma$, so each summand of \eqref{eq:series_expansion_of_expected_value} must vanish for $n>0$,  for all   $\rho_0$ and  $O_{1,3}\in\mathfrak{A}(\mathcal{R}_{1,3})$. The condition $[ \mathcal{E}_{O_2,\mathcal{B}}(O_3),O_1]=0$  looks to be sufficient to ensure no-signalling, as one expects $[ \mathcal{E}_{O_2,\mathcal{B}}(O_3),O_1]=0$ to imply $[ \mathcal{E}_{O_2,\mathcal{B}}(O_3),O_1]_n=0$ for $n>1$. This is the case if we restrict our attention to operator algebras $\mathfrak{A}$ that are bounded and densely defined on $\mathcal{H}$, as we shall assume for the rest of this argument\footnote{This can be seen by using the bounded linear transform theorem~\cite{reed1980functional}.}. Note that such a class of algebras includes the type III von Neumann algebras  underpinning algebraic QFT. 

Under the same assumptions, the vanishing of the commutator is also a necessary condition for no-signalling.  Denoting $[ \mathcal{E}_{O_2,\mathcal{B}}(O_3),O_1]$ by $O$ and using  \eqref{eq:series_expansion_of_expected_value} at order $n=1$, no-signalling implies 
$\bra{\psi}O\ket{\psi}  = 0$,
for all  physical pure states $\rho_0=\ket{\psi}\bra{\psi}$. Since $iO$ is self-adjoint,  this implies  $O\ket{\psi} = 0$ for all $\ket{\psi}$: for any element of an orthogonal  basis set $\{\ket{e_i}\}$ we have $\bra{e_i}O\ket{e_i}=0$. Consider  $\ket{\zeta} = a \ket{e_i}+ b \ket{e_j}$ and $\ket{\zeta'} = a \ket{e_i}+ ib \ket{e_j}$, where $a,b\in\mathds{R}$. Then  $\bra{\zeta^{(\prime)}} iO \ket{\zeta^{(\prime)}} =0$  implies 
$ab\bra{e_i}O\ket{e_j}=0$ for all  $a,b\in\mathds{R}$ and, hence,  all matrix elements of $O$ vanish, thus  establishing our claim. 

It is now clear why operators $O_2$ generically signal. While $[O_1,O_3]=0$, in general
$[O_{1,3},O_2]\neq 0$. Since the projectors $E_n$ associated to $O_2$, for any choice of $\mathcal{B}$, belong to  $\mathfrak{A}(\mathcal{R}_2)$, typically  $[E_n , O_{1,3}]\neq 0$, so it seems plausible that $\left[ \mathcal{E}_{O_2 ,\mathcal{B}}(O_3),O_1\right]  \neq 0$ would be reasonably generic. Of course, it might  be the case that $[E_n , O_{1,3}]\neq 0$ for all $n\in\mathcal{I}$, while  $\left[ \mathcal{E}_{O_2 ,\mathcal{B}}(O_3),O_1\right] = 0$. Hence, it is  important to investigate specific operators $O_2$ and resolutions $\mathcal{B}$. In doing so, we encounter seemingly innocuous operators that  signal, as we shall describe below.

It is worth noting that the above claim applies more generally to state update maps that are not necessarily ideal measurements. Specifically, one could replace the ideal measurement update map, $\mathcal{E}_{O_2, \mathcal{B}}$, in above claim/argument by any other state update map, $\mathcal{E}$, for which $\langle O_3 \rangle$ remains well defined. This includes POVM's and the probe prescription in~\cite{Fewster:2018qbm,bostelmann2020impossible}. Furthermore, the fact that the probe prescription in~\cite{Fewster:2018qbm,bostelmann2020impossible} does not signal is consistent with our criterion in the sense that the probe update map, $\mathcal{E}_P$, satisfies the above condition: $[ \mathcal{E}_P(O_3) , O_1 ] =0$ for all $O_{1,3}\in\mathfrak{A}(\mathcal{R}_{1,3})$.

Applying our condition to state update maps more generally, we can go further and rephrase the causal criterion in a way that is independent of observables $O_{1,3}$ and their respective regions $\mathcal{R}_{1,3}$. First, one notices that for any pair of regions, $\mathcal{R}_1$ and $\mathcal{R}_3$, with spacelike closures, the respective sub-algebras satisfy $\mathfrak{A}(\mathcal{R}_3 ) \subseteq \mathfrak{A}(\mathcal{R}_1 )'$, where $\mathfrak{A}'$ denotes the \emph{commutant} of $\mathfrak{A}$, i.e. the space of all observables that commute with all observables in $\mathfrak{A}$. For a product Hilbert space, $\mathcal{H}_A\otimes \mathcal{H}_B$, we have $(B(\mathcal{H}_A)\otimes \mathds{1}_B)' = \mathds{1}_A\otimes B(\mathcal{H}_B)$ and $(B(\mathcal{H}_A)\otimes \mathds{1}_B)'' = B(\mathcal{H}_A)\otimes \mathds{1}_B$, where $B(\mathcal{H})$ denotes the set of bounded operators on $\mathcal{H}$.

The commutator condition for a given update $\mathcal{E}$ can now be rephrased in terms of a \emph{single} region $\mathcal{R}$ and its associated sub-algebra $\mathfrak{A}(\mathcal{R})$. Specifically, $[ \mathcal{E}(X) , Y ]=0$ for all self-adjoint $Y\in \mathfrak{A}(\mathcal{R})$ and all self-adjoint $X\in \mathfrak{A}(\mathcal{R})'$. This immediately implies that $[ \mathcal{E}(X) , Y ]=0$ for all $Y\in \mathfrak{A}(\mathcal{R})$ and all $X\in \mathfrak{A}(\mathcal{R})'$, and hence $\mathcal{E}(\mathfrak{A}(\mathcal{R})')\subseteq \mathfrak{A}(\mathcal{R})'$. For $\mathcal{E}$ to be causal this must hold for all regions $\mathcal{R}$, which leads to the intuitively sensible
\begin{claim}[]\label{claim2}
A state update, $\mathcal{E}$, will not enable a superluminal signal iff it is commutant non-increasing.
\end{claim}

\noindent By \emph{commutant non-increasing} (c.n.i.) we mean that $\mathcal{E}(\mathfrak{A}(R)')\subseteq \mathfrak{A}(R)'$ for all regions $\mathcal{R}$. Clearly the c.n.i. property is also sufficient: if $\mathcal{E}(X) \in \mathfrak{A}(\mathcal{R})'$, then $\mathcal{E}(X)$ commutes with every $Y\in\mathfrak{A}(\mathcal{R})$, implying our no-signalling condition. For a more in-depth analysis of an update map's effect on spacetime subalgebras see~\cite{Jubb2021}.


\section{Examples} 

Focusing again on observables and their associated ideal measurements, let us consider some simple examples where $\mathcal{H}=\mathcal{H}_A \otimes \mathcal{H}_B$, where Aoife and Beolagh have access to only $\mathcal{H}_A$ and $\mathcal{H}_B$ respectively. That is, $O_1=O_A\otimes\mathds{1}_B$ and $O_3=\mathds{1}_A\otimes O_B$.

\subsection{Direct sum of local compact observables}

We shall use the causal criterion to show that the direct sum of local observables cannot signal, in agreement with expectations. In particular, this subsumes the case of the manifestly local Bell-type experiments, which of course cannot be used to signal and are realizable in local isolated laboratories.

Specifically, for Caoimhe's measurement we consider sums of local operators
\begin{equation}
O_2 = C_A \otimes  \mathds{1}_{B} + \mathds{1}_A \otimes C_B \;\;\; ,
\end{equation}
for some compact self-adjoint operators $C_{i}$, $i=A,B$, on their respective Hilbert spaces $\mathcal{H}_{i}$. As we shall demonstrate, such operators cannot signal since the conditions of Claim \autoref{claim1} are met for all $C_{i}$.  First, we diagonalize $C_{i}$ and write it as
\begin{align}\label{eq:cAB_decomposition}
C_{i} = \sum_{n=1}\mu_{i}^{(n)} E_{i}^{(n)} \;\;\; ,
\end{align}
where the distinct eigenvalues $\mu_{i}^{(n)}$ are ordered in decreasing magnitude, and $E_{i}^{(n)}$ are the corresponding projectors (not necessarily rank-1). If $C_{i}$ has a kernel,  we  denote the corresponding projector as $E_{i}^{(0)}$ and write $\mu_{i}^{(0)}=0$. 
We can now write $O_2$ as
\begin{equation}
O_2 = \sum_{n,n'=0}\left(\mu_A^{(n)}+\mu_B^{(n')}\right)E_A^{(n)}\otimes E_B^{(n')} \;\;\; .
\end{equation}
Since $O_2$ is also compact and self-adjoint, we can  write it similarly as 
\begin{equation}
O_2 = \sum_{a=1} \sigma^{(a)}P^{(a)} \;\;\; ,
\end{equation}
where, in general, a given projector $P^{(a)}$ will be a sum of terms of the form $E_A^{(n)}\otimes E_B^{(n')}$. For every choice of $n$ and $n'$,  $E_A^{(n)}\otimes E_B^{(n')}$ will appear in one and only one $P^{(a)}$ and, more importantly, $P^{(a)}$ cannot contain two (or more) terms that share a factor $E_{i}^{(n)}$. For example, if 
\be
P^{(a)} = E_A^{(n)}\otimes E_B^{(n')}+ E_A^{(n)}\otimes E_B^{(m')} +\cdots
\ee
 for $n'\neq m'$, then
$
\mu_A^{(n)}+\mu_B^{(n')} = \mu_A^{(n)}+\mu_B^{(m')}
\Rightarrow\mu_B^{(n')} = \mu_B^{(m')}$,  in contradiction with our initial setup with distinct $\mu_i^{(n)}$.

Given the above conditions on the form of the projectors $P^{(a)}$, one can show that
\begin{align}
\mathcal{E}_{O_2}(O_3) &  = \sum_{a=0} P^{(a)} \cdot \mathds{1}_A\otimes O_B \cdot P^{(a)}  \nonumber
\\
& = \sum_{n,n'=0} E_A^{(n)} \otimes \left( E_B^{(n')}O_B E_B^{(n')} \right) \nonumber
\\
& = \mathds{1}_A \otimes \mathcal{E}_{C_B}(O_B) \;\;\; ,
\end{align}
where  we have omitted the resolution $\mathcal{B}$ as  $O_2$ is compact self-adjoint, and we assume our measurement apparatus perfectly resolves the eigenvalues. Finally, given $O_1= O_A \otimes \mathds{1}_B$, it follows that $\left[ \mathcal{E}_{O_2 }(O_3),O_1\right]  = 0$, and hence there can be no signal. From the last line above we also see that $\mathcal{E}_{O_2}$ is commutant non-increasing, as it outputs an operator that is trivial on $\mathcal{H}_A$.

This non-signalling nature of sums of local observables agrees with the expectation that  smearings of local field operators $\phi(x)$ over  (subsets of)   spacelike hypersurfaces, will not signal \cite{Benincasa:2012rb}. It does not however address the physically relevant case (in that the operators are well-defined) of smearing over spacetime subregions of $\mathcal{M}$, as we shall discuss in our concluding remarks.

\subsection{Tensor product of local compact observables}

Let us now turn to an example that does signal:  $O_2 = C_A\otimes C_B$. This  contradicts the standard expectation that measurements of operators of the form $A\otimes B$ are locally realizable  \cite{Beckman:2001qs}. As we shall see, the signal requires that $O_2$ be degenerate, which may explain why it was not observed in \cite{Beckman:2001qs}, where the observables were assumed to be non-degenerate.

Using the decompositions in~\eqref{eq:cAB_decomposition} we can write $O_2=C_A\otimes C_B$ as
\be
O_2  = \sum_{a=1}  \sigma^{(a)}P^{(a)}
= \sum_{n,n'=1}\mu_A^{(n)} \mu_B^{(n')} E_A^{(n)}\otimes E_B^{(n')} \;\;\; .
\ee
For any $a>0$ we have the same conditions on the projectors $P^{(a)}$ that we had in the previous example.  However, if $C_{i}$ have non-trivial  kernels, then there is an important difference:  $E_A^{(0)}\otimes E_B^{(n)}$ and  $E_A^{(n)}\otimes E_B^{(0)}$  project a state into the kernel of $O_2$, and hence  $P^{(0)}$ is given by
\begin{equation}
P^{(0)} = E_A^{(0)}\otimes \mathds{1}_B +E_A^{(0)}{}^{\perp} \otimes E_B^{(0)} \;\;\; ,
\end{equation}
where $E_i^{(0)}{}^{\perp}=\mathds{1}_i - E_i^{(0)}$. We now have that
\begin{align}\label{eq:O2_on_O3_product_case}
\mathcal{E}_{O_2}(O_3)   = & E_A^{(0)}\otimes O_B + E_A^{(0)}{}^{\perp}\otimes E_B^{(0)} O_B E_B^{(0)}  \nonumber
\\
& + \sum_{n,n'=1}E_A^{(n)}\otimes E_B^{(n')} O_B E_B^{(n')}  \nonumber
\\
= & E_A^{(0)}\otimes O_B + E_A^{(0)}{}^{\perp} \otimes \mathcal{E}_{C_B}( O_B)  \; ,
\end{align}
and, hence,
\begin{equation}
\left[ \mathcal{E}_{O_2 }(O_3),O_1\right] = \left[ E_A^{(0)} , O_A \right]\otimes \left( O_B -  \mathcal{E}_{C_B}( O_B) \right) \; .
\end{equation}
If \textit{i)} $C_A$ has a kernel projector, $E_A^{(0)}$, which does not commute with $O_A$, and if \textit{ii)} the non-selective map $\mathcal{E}_{C_B}$ has a non-trivial action on $O_B$\footnote{For instance, if $[C_B,O_B]=0$ (which is the case if $C_B$ and $O_B$ belong to subalgebras of mutually spacelike regions) then $\mathcal{E}_{C_B}( O_B)=O_B$.}, then there will be a signal. The last line of~\eqref{eq:O2_on_O3_product_case} can be rewritten as
\begin{align}
\mathcal{E}_{O_2}(O_3) = & \mathds{1}_A\otimes \mathcal{E}_{C_B}( O_B) 
\nonumber
\\
& +E_A^{(0)}\otimes\left( O_B -  \mathcal{E}_{C_B}( O_B) \right) \; ,
\end{align}
which shows that $\mathcal{E}_{O_2}(O_3)$ is only in the commutant of $B(\mathcal{H}_A)\otimes \mathds{1}_B$ if \textit{i)} or \textit{ii)} is false. Note, the above derivation can also be used in the case where $C_i$ has no kernel by multiplying every $E_{i}^{(0)}$ by $0$.

These rather generic conditions are easily met, as demonstrated by the following  simple two-qubit example. Let the initial state be $\rho_0 = \ket{\psi}\bra{\psi}$, where 
\begin{equation}
\ket{\psi} = \ket{0} \otimes \frac{1}{\sqrt{2}}  \left( \ket{0}+\ket{1} \right) \;\;\; .
\end{equation}
Aoife  kicks $\rho_0$ with the operator $U_\gamma = e^{i \gamma O_1}$, where $O_1 = \sigma^{x} \otimes \mathds{1}$ \footnote{Here $\sigma^{x,y,z}$ denote the usual Pauli matrices.}. Next, Caoimhe measures 
\be
O_2 = \ket{1}\bra{1} \otimes \sigma^z \;\;\; ,
\ee
 which has the kernel projector $P^{(0)}=\ket{0}\bra{0} \otimes \mathds{1}$, and two other projectors $P^{(1)} = \ket{1}\bra{1}\otimes \ket{0}\bra{0}$ and $P^{(2)} = \ket{1}\bra{1}\otimes \ket{1}\bra{1}$ corresponding to the eigenvalues $+1$ and $-1$ respectively. Note, Caoimhe's observable is rather pedestrian; it is nothing out of the ordinary being a separable tensor product of observables on the constituent systems.
 
   Finally, Beolagh measures the expectation value of $O_3 = \mathds{1} \otimes \sigma^x$.  One can verify that
\begin{equation}
\left\langle O_3 \right\rangle = \cos^2 (\gamma) \;\;\; .
\end{equation}
Since the expectation value depends on $\gamma$, Aoife can signal to Beolagh.

It may seem counterintuitive that an ideal measurement of a \emph{separable} operator $O_2$ on a \emph{separable} state enables a signal. To gain some intuition for why this is so, we can ``realize'' the ideal measurement of $O_2$ via a causally connected two stage protocol, specifically a Local Operations and Classical Communications (LOCC) protocol: Caoimhe first measures the $z$-spin on qubit $A$. If the spin on qubit $A$ is down, then Caoimhe does nothing on qubit $B$. If the spin on $A$ is up, then Caoimhe measures the $z$-spin on qubit $B$. One can verify that this LOCC protocol amounts to the map $\mathcal{E}_{O_2}(\cdot)$. Clearly, this LOCC protocol requires a signal to propagate from qubit $A$ to $B$, as Caoimhe's operation on qubit $B$ is conditioned on the result on qubit $A$.

Given that this LOCC realization of an ideal measurement of $O_2$ exists, one might be skeptical that an experimenter could implement an ideal measurement of $O_2$ faster than the light travel time from $A$ to $B$. An optimistic experimenter, on the other hand, may still be hopeful that some faster realization of the measurement exists. The fact that $\mathcal{E}_{O_2}(\cdot )$ fails our criterion, however, illustrates that any such realization is impossible, as it would enable a superluminal signal from $A$ to $B$.

\section{Conclusions} The appropriate  context for quantum causal considerations is relativistic QFT, which  elegantly captures the notions of locality and causality in its very foundations \cite{Haag:1992hx}. This is not in dispute.  What the present and previous related \cite{Sorkin:1993gg, Popescu:1993xc, Beckman:2001qs, Benincasa:2012rb} results imply is that the most general, logically consistent, application of the quantum measurement framework to observables in spacetime, as given in \cite{Sorkin:1993gg}, is problematic, whether it be in QFT or otherwise.

There are a variety of attitudes one could take. First would be to place causal constraints on the class of observables that may be measured \cite{Popescu:1993xc, Beckman:2001qs}. In this case, as well as various non-separable observables \cite{Sorkin:1993gg, Popescu:1993xc, Beckman:2001qs, Benincasa:2012rb}, we have lost (a subset of) observables of the form $A\otimes B$, with the obvious generalization to multipartite systems.  Local sums, $A\otimes \mathds{1}+ \mathds{1}\otimes B$, are causality respecting, at least for compact $A$ and $B$. This supports the expectation that local field operators smeared over spacelike hypersurfaces will not signal. However, the objects of algebraic QFT are (un)bounded operators in open subsets of space\emph{time}, which may signal yet in an interacting theory. Consider a scalar field $\phi(x)$ smeared against a test function with bounded spacetime support in some region $\mathcal{R}$. This operator can be mapped back to an operator, $O$, on the  intersection of any spacelike hypersurface and the causal past of  $\mathcal{R}$. In a free theory, $\phi(x)$ at $x$ can be written as a linear sum of field operators on the  intersection of any spacelike hypersurface and the causal past of $x$.  By linearity, $O$ is also a linear sum of field operators and, naively, we should not expect any signalling. However, in an interacting theory this fails, $O$ will not generically be a linear sum of field operators and, in light of our second example, this is suggestive of signalling. Establishing this possibility is a technically challenging question and will be treated elsewhere. Since the  projection postulate and no-signalling criterion can be applied to the operators of type III von Neumann algebras, this impinges on the question of measurement in algebraic QFT. 

This motivates the second possibility: to reassess the applicability of the standard measurement prescription to QFT, or at least to reevaluate how to apply the projection postulate. The best known proposal, that the projection takes place on the past lightcone of the measurement region \cite{Hellwig:1970st}, certainly does not address the issue of signalling.

It has been noted that restricting the use of the projection postulate to measurement regions that are totally timelike/spacelike related rules out signalling by fiat \cite{Sorkin:1993gg}, but this can break down in spacetimes where the spatial slices (surfaces of constant time) are compact. More specifically, consider an ultrastatic spacetime\footnote{In an ultrastatic spacetime the metric is of the form: \newline $g=-dt^2 + h_{ij}dx^i dx^j$, i.e. a fixed spatial geometry ``cross'' time.} with compact spatial slices, and some measurement region, $\mathcal{R}$, that is small enough to not contain an entire spatial slice. The \emph{total future} of $\mathcal{R}$, i.e. the set of spacetime points that are to the future of \emph{every} point of $\mathcal{R}$, contains an entire spatial slice. In the Heisenberg (or algebraic QFT) picture it is then clear that the subalgebra of observables associated with the total future of $\mathcal{R}$ is in fact the entire algebra of all observables. The seemingly safe position of adopting the projection postulate for only those observables in the total future of $\mathcal{R}$ amounts, in this case, to adopting it for \emph{all} observables, including the troubling cases matching the setup in figure~\ref{fig:setup}.

Even if one assumes a spatially non-compact universe, one is still compelled to give an account of what happens for sets of measurements that are not totally timelike/spacelike related. For example, how are we to describe the situation in which a second measurement region \emph{partially} intersects the future of the first measurement region? There is no \textit{a priori} reason to expect that such a pair of experiments cannot be done in nature.

Leaving these subtleties aside, constructing a complete and causality respecting measurement model of QFT is highly non-trivial. There are, for example, causal QFT measurement models in the spirit of von Neumann \cite{Fewster:2018qbm,bostelmann2020impossible, Doplicher:2009xh, Grimmer:2019ahl}. The probe prescription~\cite{Fewster:2018qbm,bostelmann2020impossible} in particular offers a physically motivated causal measurement model, whereby a probe quantum field is locally coupled to the main quantum field of interest. The probe field is measured and the result is interpreted as a measurement of the main field. After this measurement the probe field is discarded, meaning that no further observables of the probe field can be measured. If one wishes to make subsequent measurements of the main field, one must further couple (distinct) probe fields and measure them. The state update maps arising in this prescription (by tracing out the probe fields) satisfy the no-signalling condition due to the locality of the coupling in the setup.

On the other hand, in the spirit of von Neumann, this model also pushes back the problem of signalling to the probe observables. If one wishes to discuss subsequent measurements on the \emph{probe field}, in a way that is non-signalling, one must further introduce a probe for the probe, and so on \textit{ad infinitum}. This chain can of course be halted if one restricts the set of allowed probe observables \textit{\`a la} the above discussion. In~\cite{Fewster:2018qbm,bostelmann2020impossible} it is halted by construction, since the abandoning of the probe field post-measurement means we cannot measure any of its observables again. The act of discarding the probe is typical of measurement models in non-relativistic quantum theory, but it is more suspect when the probe is itself a quantum field, especially if the field is considered to be sufficiently fundamental. For example, if we make a single measurement of an electron's position, using light as a probe, we do not want to prohibit subsequent measurements of observables of the electromagnetic quantum field.

Nonetheless, the probe prescription generates a concrete class of physically motivated, and causal, state update maps for QFT. In contrast, our criterion --- which poses a theoretical limit for allowed state update maps --- can currently only rule out measurements of observables in an \emph{ad hoc} manner. An interesting question, then, is whether the theoretical limit matches the probe state update maps. In other words, does every causal state update map arise from some locally coupled probe setup?

An alternative viewpoint to all of the above is to interpret signalling as a causal constraint on the \emph{measurement resolution} achievable within a given spacetime region.  For any operator $O_2$ one can always coarse-grain the measurement resolution $\mathcal{B}$ such that the signal is killed. For example, if the measurement apparatus for our two-qubit example, where $O_2= \ket{1}\bra{1} \otimes \sigma^z$, is not able to distinguish the $\pm 1$ eigenvalues, signalling is not  possible. It is also interesting to note that the realization of Caoimhe's measurement as a LOCC protocol reduces, in this case, to a single measurement of the z-spin of qubit $A$, meaning that no signal needs to propagate from $A$ to $B$.

The final possibility is to couch the foundations of QFT and measurement entirely within the manifestly  relativistic framework of the sum-over-histories approach pioneered by Dirac and Feynman, forgoing the complementary picture of Hilbert space, operators and transformation theory.   This perspective has been advocated for independent  reasons, that   nonetheless are ultimately  related to the issue of signalling \cite{Hartle:1991bb, Hartle:1992as, Hartle:1993ip, Sorkin:1994dt, Sorkin:2010kg}. 

\acknowledgments 
We are grateful for stimulating conversations with Aiyalam P.~Balachandran,  Chris Fewster, Jorma Louko, Denjoe O'Connor, Rafael Sorkin and especially to Fay Dowker.  LB is supported by a Schr\"odinger Fellowship. IJ is supported by an Irish Research Council Fellowship (GOIPD/2018/180). GK is supported by a Schr\"odinger Fellowship and acknowledges support from Science Foundation Ireland through Career Development Award 15/CDA/3240.

\end{document}